\documentclass[11pt]{article}
\usepackage{epsf}

\def\1ad{\mbox{\normalsize $^1$}}
\def\2ad{\mbox{\normalsize $^2$}}
\def\3ad{\mbox{\normalsize $^3$}}
\def\4ad{\mbox{\normalsize $^4$}}
\def\5ad{\mbox{\normalsize $^5$}}
\def\6ad{\mbox{\normalsize $^6$}}
\def\7ad{\mbox{\normalsize $^7$}}
\def\8ad{\mbox{\normalsize $^8$}}

\def\npb#1#2#3{{\it Nucl. Phys.} {\bf B#1} (#2) #3 }
\def\plb#1#2#3{{\it Phys. Lett.} {\bf B#1} (#2) #3 }

\def\ijmpa#1#2#3{{\it Int. J. Mod. Phys.} {\bf A#1} (#2) #3 }

\def\bb#1{{\tt hep-th/#1}}

\def\jhep#1#2#3{{\it JHEP} {\bf #1} (#2) #3 }
\def\atmp#1#2#3{{\it Adv. Theor. Math. Phys.} {\bf #1} (#2) #3 }
\def\bb#1{{\tt hep-th/#1}}

\def\dj{\hbox{d\kern-0.347em \vrule width 0.3em height 1.252ex depth
-1.21ex \kern 0.051em}}

\def\half{{1\over 2}\,}

\def\Tr{{\rm Tr\,}}

\newcommand{\leqsim}{\,\raisebox{-0.6ex}{$\buildrel < \over \sim$}\,}
\newcommand{\geqsim}{\,\raisebox{-0.6ex}{$\buildrel > \over \sim$}\,}
\newcommand{\be}{\begin{equation}}
\newcommand{\ee}{\end{equation}}
\newcommand{\ba}{\begin{eqnarray}}
\newcommand{\ea}{\end{eqnarray}}

\newcommand{\ie}{\mbox{{\em i.e.~}}}

\def\half{\frac{1}{2}}

\newcommand{\ben}{\begin{equation*}}
\newcommand{\een}{\end{equation*}}
\newcommand{\ban}{\begin{eqnarray*}}
\newcommand{\ean}{\end{eqnarray*}}
\newcommand{\brr}{\begin{array}}
\newcommand{\err}{\end{array}}
\newcommand{\bc}{\begin{center}}
\newcommand{\ec}{\end{center}}

\begin{document}

\newcommand{\sheptitle}
{Stringy Fuzziness as the Custodian of Time-Space Noncommutativity}
\newcommand{\shepauthora}
{{\sc
J.L.F.~Barb\'on}
\footnote[1]{On leave from 
Departamento de F\'{\i}sica
de Part\'{\i}culas. Universidad de Santiago de Compostela, Spain.}
}

\newcommand{\shepaddressa}
{\sl
Theory Division, CERN \\ 
 CH-1211 Geneva 23, Switzerland \\
{\tt barbon@mail.cern.ch}}

\newcommand{\shepauthorb}
{\sc
E.~Rabinovici}

\newcommand{\shepaddressb}
{\sl
Racah Institute of Physics, The Hebrew University \\ Jerusalem 91904, Israel
 \\
{\tt eliezer@vms.huji.ac.il}}

\newcommand{\shepabstract}
{
We study aspects of obtaining field theories with noncommuting time-space
coordinates as limits of open-string theories in constant electric-field
backgrounds. We find that, within the standard closed-string backgrounds, 
there is an obstruction to 
 decoupling the time-space noncommutativity scale from
that of the string fuzziness scale. We speculate that
 this censorship may be string-theory's 
 way of protecting the causality and unitarity structure.
We study the moduli space of the obstruction in terms of the open-
and closed-string backgrounds. Cases of both zero and infinite brane
tensions as well as zero string couplings are obtained.
 A decoupling can be achieved formally by considering complex values
of the dilaton and inverting the role of space and time in the
light cone. This is reminiscent of a black-hole horizon.
 We study the corresponding supergravity solution
in the large-$N$ limit and find that  the geometry has a naked
singularity at the physical scale of noncommutativity.
}

\begin{titlepage}
\begin{flushright}
CERN-TH/2000-133\\
RI/00-05-02\\
{\tt hep-th/0005073}\\

\end{flushright}
\vspace{0.5in}
\begin{center}
{\large{\bf \sheptitle}}
\bigskip\bigskip \\ \shepauthora \\ \mbox{} \\ {\it \shepaddressa} \\
\bigskip\bigskip  \shepauthorb \\ \mbox{} \\ {\it \shepaddressb} \\
\vspace{0.5in}

{\bf Abstract} \bigskip \end{center} \setcounter{page}{0}
\shepabstract
\vspace{0.5in}
\begin{flushleft}
May 2000
\end{flushleft}


\end{titlepage}

\newpage

%
%

\section{\sf Introduction}
\setcounter{equation}{0}

The list of properties of manifolds which are not ambiguous when studied
by point particles, but can be ambiguous when probed by strings, has
lengthened considerably over the years. The list includes  the
geometric data (such as the metric), the topology and the dimensionality
of the manifold. There are by now many examples where this occurs for both
large and small manifolds. Recently, there has been a new addition to this
list of potentially ambiguous properties: theories on noncommutative 
manifolds \cite{conn} 
were found to be equivalent \cite{SW} to other theories on commutative
manifolds. These theories had a Dirac--Born--Infeld limit and thus contained 
many derivatives. When viewed as effective theories for open-string dynamics,
it seems that the end-points of the open strings 
are responsible for  such noncommutative properties of the space-time
coordinates:
\be
\label{ncc}
[x^\mu , x^\nu ] = i\,\theta^{\mu\nu}
.\ee
 The
main emphasis was on theories for which only spatial 
coordinates of the manifold were noncommuting, \ie $\theta^{0i} =0$ in 
(\ref{ncc}).  
The framework of string
theory in the presence of large magnetic-field backgrounds was
very useful
for such an analysis \cite{cds, dhull, bunch}.
 Such a theory seems to contain two explicit scales: 
the string scale $\sqrt{\alpha'}$ 
 (which we should also be permitted  to hope, will  
eventually  be found to be spontanously generated) and the noncommuting
length scale $\sqrt{|\theta|}$.

Seiberg and Witten have found a limit of moduli space in which 
 the two scales can be decoupled 
 \cite{SW}. This limit is a field theory decoupled from
string oscillators on a noncommutative manifold. It is interesting to
dare 
string theory with an extra challenge, that is to try and obtain out of
it a field-theory limit that would realize time-space noncommutativity
while decoupled from the string scale. We will find that is not possible,
at least within standard sigma-model backgrounds.

It seems clear that the time-space noncommutativity requires a reevaluation  
of the  role of the Hamiltonian, of causality and unitarity.   
 On the other hand, string theory has been known in the past to
take care of such issues. For example, the fact that D-branes cannot travel 
at a speed larger than the speed of light reflects itself, in a T-dual
picture \cite{bachas},
 in the existence of a bound on the possible strength of an 
electric-field background \cite{bur}.
 Another recent example is the realization in
ref. \cite{teleo} that standard open strings in flat Minkowski space 
show the same paradoxical features expected from a time-space noncommutative
theory, most notably an apparently acausal  
 behaviour in scattering experiments. 

Indeed, there
 is a time-space uncertainty principle (see \cite{yone} for a review
and a collection of references),
 derived on heuristic grounds,
 which is valid in principle for {\it both} open- and
closed-string theories:
\be
\label{strun}
|\Delta t \,\Delta x |  \geqsim \alpha'
.\ee
It operates at the string scale and should apply
independently of the background, provided it is sufficiently smooth on
the string scale.

On the other hand, the peculiar noncommutativity properties we are
interested in are specifically associated to open-string end-points \cite{
dipoles}. Therefore,
it would be very interesting to disentangle these effects from whatever
is masked  by `standard' stringy fuzziness in (\ref{strun}).

 Following the strategy of the purely spatial noncommutative
examples,  taking the Seiberg--Witten limit
of the
  corresponding noncommutative string backgrounds,      
 one finds an obstruction of the same nature as the maximal
electric field. It is not possible to decouple the string fuzziness from
the time-space noncommutativity. It is as if the string fuzziness served 
as a custodian of causality and unitarity and, as such, could not be decoupled
from the noncommutativity scale.

The structure of the paper is as follows:
in section 2 we encounter a maximal electric-field obstruction to
decoupling the time-space noncommutativity and the
string scale. We analyse the moduli space of such backgrounds in terms 
of both open-
and closed-strings geometrical data. We analyse properties of these two 
families of backgrounds and find that the string theory cannot be decoupled from
the field theory.  This leads us to suggest in section 3 
that 
time-space 
noncommutativity is `censored'  by the string fuzziness. We further explore this
notion  in section 4, where we do a formal continuation of the string
data in such a way as to  reproduce the perturbative expansion of
a time-space noncommutative theory.  
This involves imaginary dilatons and inverted space-time light-cone
coordinates, not so dissimilar from the inversion occurring  near the horizon
of a black hole. 
 A dual
supergravity master field in the
appropriate large-$N$ strong-coupling limit is obtained. It has a naked
singularity, whose possible resolution by stringy effects    
might reinstate the `custodial effect'.

\section{\sf Two Families of Critical Field Singularities}

The classical open-string dynamics in a background with metric $g_{\mu\nu} $
 and  NS $B_{\mu\nu}$-fields 
is controlled by the open-string metric $G_{\mu\nu}$ and the noncommutativity
parameter $\theta^{\mu\nu}$, which determine the world-sheet propagator. At
the tree level: 
\be
\label{propd}
\left\langle X^\mu (\tau) \,X^\nu (\tau')\right\rangle_{\rm disk} =
 -\alpha' \,G^{\mu\nu} \,
{\rm log}\,(\tau-\tau')^2 + {i\over 2} \,\theta^{\mu\nu}\,\epsilon (\tau-\tau')
.\ee
The open-string parameters
 are related to the sigma-model fields $g_{\mu\nu}, B_{\mu\nu}$ by  
\be
\label{mapnc}
G^{\mu\nu} + {\theta^{\mu\nu} \over 2\pi\alpha'} = \left({1\over
g + 2\pi\alpha' B}\right)^{\mu\nu}
,\ee
with $\theta^{\mu\nu}$ entering the commutation relations of the string
zero modes as in (\ref{ncc}). Therefore, 
 the noncommutativity properties of the open-string theory are associated
to non-vanishing  
NS $B$-field backgrounds.  In particular, in order to induce time-space
noncommutativity $\theta^{0i} \neq 0$, we need non-vanishing
 electric components of
the NS $B$-field. 
 
For magnetic $B$-field backgrounds, $B_{0i} = 0=\theta^{0j}$, 
 one can expose the noncommutative
properties  by explicitly disentangling the $\theta$-nonlocality from
the stringy fuzziness.  
Such a decoupling was studied 
in full detail by Seiberg and Witten in \cite{SW}.  It involves a zero-slope
limit of the string theory $\alpha' \rightarrow 0$, holding fixed
the open-string metric $G_{\mu\nu}$ and the open-string NC parameter
$\theta^{\mu\nu}$, resulting in a low-energy noncommutative Yang--Mills
theory, with interactions specified in terms of Moyal products. In this
limit
\be
\label{nstheta}
\theta^{\mu\nu} \rightarrow \left({1\over B}\right)^{\mu\nu}.
\ee
 It is    
also found necessary to scale the string coupling so that  
\be
\label{effgs}
G_s = g_s \left({-{\rm det}(g+2\pi\alpha' B) \over -{\rm det}(g)}\right)^{\half}
\ee
is the effective string-loop expansion parameter in the limit. 
The zero-slope limit involves scaling the sigma-model metric to zero
as $g_{\mu\nu} \sim (\alpha')^2$ and leaving $B_{\mu\nu}$ fixed. Therefore,
there seems to be a clear obstruction when the $B$-field has electric
components. In that case, the matrix
$$
g+b \equiv g+2\pi\alpha' B 
$$
will eventually have imaginary eigenvalues, and the open-string background
is ill-defined. To be more specific,   
 suppose that $B_{\mu\nu}$ is skew-diagonalized   and consider the time-like
$2\times 2$ block. Parametrizing this block by
\be
(g_{\mu\nu}+ 2\pi\alpha' B_{\mu\nu}) =
 \left(\begin{array}{cc} -{\bar g} & -{\bar b}
 \\ {\bar b} & {\bar g}\end{array}
\right), 
\qquad (G^{\mu\nu})\equiv
 \left(\begin{array}{cc} -{\bar G}^{-1} & 0 \\ 0 & {\bar G}^{-1}\end{array}
\right), 
 \qquad (\theta^{\mu\nu}) \equiv \left(\begin{array}{cc} 0 & {\bar \theta} \\ -{\bar \theta} 
 & 0\end{array}
\right)
,\ee
one finds the relations
\be
\label{directrel}
{\bar G}^{-1} = {{\bar g} 
 \over {\bar g}^2 - {\bar b}^2}, \qquad {\bar \theta} =
 -2\pi\alpha' {{\bar b} \over {\bar g}^2-{\bar b}^2}
\ee
and
\be
\label{effstr}
G_s = g_s \sqrt{{\bar g}^2 - {\bar b}^2 \over {\bar g}^2}
.\ee
Therefore, any zero-slope limit in which the $B$-field dominates over the
metric,  $g+b \approx b$,  leads to a positive determinant
${\rm det}(g+b) >0$ and to an imaginary effective string coupling 
(\ref{effstr}).
In particular, at the vanishing locus of the determinant, ${\rm det}(g+b)=0$,
 the open-string inverse metric and noncommutative
 parameters are infinite!                   

We find that the moduli space is  divided into
 two branches: one with ${\rm det}(g+b)  
<0$, continuously connected to the fully
commutative background, and one with ${\rm det}(g+b) >0$, which seems
to be ill-defined. The critical line is characterized by a singularity
of the open-string parameters $G, \theta$.

Open strings in electric-field backgrounds are well-known to exhibit
a classical singularity at a critical value of the electric field \cite{bur,
igor}. 
This singularity can be spotted in the low-energy effective dynamics  
as the vanishing of the Dirac--Born--Infeld (DBI) Lagrangian
\be
\label{bivan}
{\cal L}_{\rm DBI} \sim {1\over g_s} \sqrt{-{\rm det}(g+b)} =0
\ee
 or, in T-dual language,
as the limiting value of the speed of light for the corresponding T-dual
D-brane \cite{bachas}.  
Thus, the critical line of singularities coincides with the classical
singularity of the DBI action.  The physical interpretation is that
the D-brane becomes effectively tensionless, and in fact tachyonic for 
${\rm det}(g+b) >0$. 

Actually, the DBI singularity at the locus ${\rm det}(g+b)=0$ does not
exhaust all the singularities in the mapping (\ref{directrel}), for it
is not well-suited for  studying the limits in the $(g,b)$ moduli space  
where both $g$ and $b$ diverge. One can factor the volume form 
                and study the submanifold  
\be
\label{othersing}
{\rm det}(1+g^{-1}\,b) =0
,\ee
which includes extra singularities at $g\sim \infty$. In fact, there is
a whole moduli space of them. Solving for the inverse relations in
(\ref{directrel}):  
\be
\label{inverserel}
{\bar g}={{\bar G}\over 1-\left({{\bar G}
{\bar \theta}\over 2\pi\alpha'}\right)^2} , \qquad {\bar b}= -{{\bar G} \left({
{\bar G}{\bar \theta}\over
2\pi\alpha'}\right) \over 1 - \left({{\bar G}{\bar \theta} \over 2\pi\alpha'}
\right)^2}
,\ee
we find that there is a family of singularities at ${\bar G} {\bar \theta} =
2\pi\alpha'$, which are solutions of (\ref{othersing}) and correspond to
divergent sigma-model backgrounds. We shall denote these singularities
as `sigma-model singularities', $g$-singularities for short,
 on account of the fact that $g,b$ diverge
at finite values of the open-string parameters $G,\theta$. The previously
identified singularities at finite $g,b$ will be  referred to as
`open-string singularities', or $G$-singularities for short,
 since there it is $G^{-1}$ and  $\theta$ that diverge.     

The effective loop-expansion parameter $G_s$  
 vanishes at both types of critical points.
 For the $G$-singularity this is obvious from
the definition in (\ref{effstr}). For the $g$-singular points we find
\be
\label{gstrsin}
G_s = g_s \sqrt{ 1 - \left({{\bar G}{\bar \theta} \over 2\pi\alpha'}
\right)^2}
.\ee
Therefore,  a classical theory
 (no string-loop corrections) is expected  at the singularities, unless
$g_s$ is scaled so as to compensate for the vanishing of the
determinant factors. 

In spite of this, the perturbative
 physics (at fixed and small string coupling $g_s \ll 1$)
 near these two families of singularities is
rather different. First, we have noticed that the $G$-singularities  
are characterized by the vanishing of the effective D-brane tension. On
the other hand, at the $g$-singularities, we have the opposite behaviour, 
a divergent D-brane tension:     
\be
\label{biblow}
\sqrt{-{\rm det}(g+2\pi\alpha' B)} = \sqrt{{\bar g}^2-{\bar b}^2} =
 \sqrt{{\bar G}^2 \over
 1 - \left({{\bar G}{\bar \theta} \over 2\pi\alpha'}
\right)^2}
 =
\sqrt{-{\rm det}(G) \over  1 - \left({{\bar G}{\bar \theta} \over 2\pi\alpha'}
\right)^2}
.\ee 
This is consistent with the idea that 
  $g$-singularities are less harmful than $G$-singularities,
at least when considering open-string dynamics. Open-string perturbation
theory is a weak-field expansion in the background of a D-brane with
some fluxes. If the effective tension of the brane vanishes, the fluctuations
are too violent and nonlinear effects cannot be controlled in perturbation
theory. We shall confirm this picture below when we study the scaling
of the general perturbative amplitude at a $G$-singular point.
 
The differences also show up at the level of the free spectrum. 
 This can be studied by direct quantization of the free
open string in background fields \cite{callan, callanlov, bur}.
 A shortcut to the answer can be
obtained from a reinterpretation of the one-loop vacuum amplitude,
\ie the annulus \cite{callanlov}: 
\be
{\cal A} = Z[g=\eta, b=0] \;\sqrt{-{\rm det}(g)} \cdot 
 \left({-{\rm det}(g+b) \over -{\rm det} (g)}\right)
\ee
with  $\eta$ the Minkowski metric,   
and
\be
Z[g=\eta, b=0] = \half \int_0^\infty {dt\over t} \,\Tr \,
 e^{-t\alpha' (p_\mu \eta^{\mu\nu} p_\nu +M^2)}
\ee
the vacuum Minkowski amplitude. 
  In terms of the open-string metric, 
\be
G= (g+b) \,{1\over g} \,(g+b) 
,\ee
we can use the identities 
\be
\label{identities}
{{\rm det}(G) \over {\rm det}(g+b)} = {{\rm det}(g+b) \over {\rm det}(g)} =
\left({{\rm det}(G) \over {\rm det}(g)}\right)^\half
\ee
to rewrite the complete 
 expression in the   form of the modified partition function: 
\be
{\cal A} =   \half \int_0^\infty {dt\over t}
 \,\Tr \, e^{-t\alpha' (p_\mu G^{\mu\nu} p_\nu +M^2
)}
.\ee
Thus,
 the effect of the background fields at the level of free-string propagation
 amounts simply to modifying the dispersion relation to 
\be
\label{disp}
G^{\mu\nu} p_\mu p_\nu + M^2 =0
.\ee
  Now, since this open-string  metric $G_{\mu\nu}$ is
singular at the $G$-singularity, we have a degenerate
dispersion relation there, whereas the free propagation of open strings
in the $G$-space-time is completely smooth at the $g$-singularities.

One can generalize the comparison between the two types of singularities
to higher-loop corrections. 
The general perturbative amplitude with $L$ open-string loops 
  comes
from a spherical worldsheet with $L+1$ holes and $n$ open-string vertex
insertions. We can determine the rough features of the 
 dependence on the background fields
by a scaling argument.

First,  
there is a factor of 
$$
\sqrt{-{\rm det}(g)} 
$$
from the translational zero modes; a factor of
$$
\left({-{\rm det}(g+b) \over -{\rm det} (g)}\right)^\half
$$
for each boundary, from the normalization of boundary states, as in
ref. \cite{callanlov} (see also \cite{andreev}). A factor
of $g_s^{L-1}$ from the dilaton; and a factor of $\lambda^n$ from the
normalization of the vertex operators, to be determined later on.
      
All together:
\be
\label{genloop}
{\cal A}_{L,n} \sim \sqrt{-{\rm det}(g)} \,\left({-{\rm det}(g+b) \over 
-{\rm det}(g)}\right)^{L+1 \over 2} \,\cdot g_s^{L-1} \cdot \lambda^n
.\ee
The normalization of the vertex operators depends on what  we 
consider as a background.
We have seen that, in the vicinity of the $G$-singularity, the
open-string metric $G_{\mu\nu}$ is singular and therefore the effective
Lagrangian is not naturally constructed as a density to be integrated
against $\sqrt{-{\rm det}(G)}$. Instead, we must use the original
sigma-model metric and write the effective action as  
\be
S_{\rm eff} = \int dx \sqrt{-{\rm det}(g)} \; {\cal L}_{\rm eff}
.\ee
From the general expression (\ref{genloop})
 we see that all vacuum amplitudes, including
the vacuum disk that determines the bare D-brane tension, vanish at the
$G$-singularity.  Regarding interactions,
 the canonical wave-function normalization of
open-string fluctuations should be fixed by requiring that the
tree-level two-point function density contains only the volume factor, \ie  
we require 
$$
{\cal A}_{0,2}  \sim \sqrt{-{\rm det}(g)} \left({-{\rm det}(g+b) \over
-{\rm det}(g)}\right)^{1 \over 2}\;g_s^{-1} \;\lambda^2 \sim
\sqrt{-{\rm det}(g)}
,$$
which determines
\be
\lambda \sim (g_s)^\half \cdot  \left({-{\rm det}(g+b) \over
-{\rm det}(g)}\right)^{-{1 \over 4}}
.\ee
Thus, in the vicinity of a $G$-singular point,
  amplitudes that are `sufficiently quantum', \ie at
high loop order ($L>-1 +n/2$), will vanish, while those that are
 `sufficiently nonlinear',
\ie with a large number of external legs ($n>2L+2$), will diverge.
Therefore,  we confirm our expectations based on the vanishing of the D-brane
effective tension at the $G$-singularities: the purely classical nonlinear
effects blow up.

 Incidentally, it is interesting to notice that a further scaling of the
string coupling $g_s \rightarrow 0$ can stabilize the effective
expansion parameter for the nonlinearities $\lambda$. In that limit
the theory becomes completely classical, albeit with complicated
nonlinear interactions, and it becomes also decoupled from closed strings.

On the other hand, at a $g$-singular point, it is the open-string metric 
that is the smooth one, and we are led to writing an effective Lagrangian
as a density with respect to the $G_{\mu\nu}$ metric. 
 Using  
 the identities (\ref{identities}) in (\ref{genloop}) we find 
\be
\label{otragen}
{\cal A}_{L,n} \sim \sqrt{-{\rm det}(G)} \;G_s^{L-1 } \;\lambda^n
.\ee
The wave-function normalization of vertex operators is now determined
with respect to the $\sqrt{-{\rm det}(G)}$ volume element:
$$
{\cal A}_{0,2} \sim \sqrt{-{\rm det}(G)} \;G_s^{-1 } \;\lambda^2
\sim \sqrt{-{\rm det}(G)}
,$$
which yields the expected
\be
\lambda \sim (G_s)^{\half}
,\ee
\ie in this case the same effective coupling governs both the nonlinearities
and the quantum corrections. Since $G_{\mu\nu}$
 is regular, the amplitudes at the $g$-singular point
scale as 
\be
\label{scale}
{\cal A}_{L,n} \rightarrow  (G_s)^{L-1 +n/2}
.\ee
This vanishes  if $L-1 +n/2 >0$. The vacuum disk (brane tension) diverges,
as we have pointed out before. From the point of view of the theory on
the brane, this is just an infinite adjustment of the vacuum energy and, unlike
the phenomenon of vanishing tension at the $G$-singular points, it should
be harmless.
 Also, the vacuum annulus amplitude is 
finite, as it corresponds to the smooth propagation of open strings in the
$G$-metric, according to the dispersion relation (\ref{disp}).  
All   interactions and loop corrections vanish.   
        
Therefore, the theory at the $g$-singular point is free. Again, we can
scale an interacting theory by further tuning $g_s$. In this case we
must take the underlying string-theory background to strong coupling. 

In summary, we have seen that the set of critical electric field singularities
has two branches, lying on the boundary of each other's moduli space. The
perturbative physics is qualitatively different at these two branches.
The most spectacular difference is the fact that the brane tension
diverges at $g$-singularities, with smooth open-string propagation, whereas
the brane becomes tensionless at the $G$-singularities, where open-string
dynamics  completely degenerates.

\section{\sf Noncommutative Censorship by Stringy Fuzziness}

We are chiefly interested in investigating under which conditions
the noncommutativity scale can be decoupled from the string scale,
\ie whether one can take a limit in which $|\theta | \gg \alpha'$,
and whether this limit is described by a noncommutative field theory
equipped with Moyal products in the time direction. 

Naively, we would like to emulate the prescription of ref. \cite{SW}
and, starting from a smooth sigma-model specified by a non-singular
pair $(g,b)=(g,2\pi\alpha' B)$ and a real $G_s$,
 we would take the limit of vanishing
metric at fixed $B$. In doing so, we have found that one hits the
wall of $G$-singularities, specified by the locus of ${\rm det}(g+b)=0$. 
In the vicinity of  this point we indeed have $|\theta| \gg \alpha'$, but
the open-string metric $G_{\mu\nu}$ degenerates, so that the
low-energy description in terms of Moyal products does not exist
along the lines of \cite{SW}. 

In fact, if we stay clear from the $G$-singular points, \ie if we
require $G_{\mu\nu}$ to be fixed and smooth, then we still hit
the $g$-singularities. From the inverse relations (\ref{inverserel}) 
we see that, starting at sub-stringy noncommutativity $|\theta| \ll \alpha'
$, any attemp at increasing it at fixed $G_{\mu\nu}$ takes us into a
$g$-singular point. At this point the noncommutativity is not decoupled
from the stringy fuzziness. Rather, it is of the same order:
\be
\label{maxnc}
|\theta|_{\rm max} = 2\pi\alpha'
,\ee
where we have normalized the fixed open-string metric to the Minkowski metric
value.   
Beyond this point, one can make $|\theta|$ larger than $\alpha'$, but
the effective coupling $G_s$ becomes imaginary and one does not
expect to find a unitary theory. 

Therefore, we see a kind of `censorship': string theory works very
hard in order to keep the length scale of time-space noncommutativity hidden
below the normal stringy fuzziness at the string length.

\section{\sf A Formal Decoupling Limit}  

Quantum field theories with time-space noncommutativity have no obvious
Hamiltonian formulation and, therefore, unitarity is a nontrivial
consistency issue for these theories. 
Since one finds difficulties in embedding {\it decoupled}
 time-space noncommutative
field theories into a well-defined string background, one is led to
suspect that these theories are truly inconsistent at the quantum level.

Still, the perturbation theory is obtained from that of purely spatial
noncommutativity in what may be
 a deceptively simple manner: we just continue a
spatial coordinate to $x^j\rightarrow ix^0$, and at the same time  also take
 $\theta^{j k}
 \rightarrow i\theta^{0k}$,
 in order to have a real $\theta^{\mu\nu}$ matrix after the rotation.   
In this process the real value of the Yang--Mills coupling does not change. 

The same manipulation takes the open-string perturbation theory,
 when written in terms of $G_{\mu\nu},\theta^{\mu\nu}$ and $G_s$, from
purely spatial to time-space noncommutativity. Namely
one performs an analytic continuation at fixed $G_s$. It is   
then natural to ask 
what this operation entails for the underlying string sigma-model, \ie the
`closed-string' parameters $g_{\mu\nu}, B_{\mu\nu}, g_s$. From the general
relations (\ref{inverserel}) and (\ref{gstrsin}),
 we see that the formal   zero-slope limit $\alpha'\rightarrow 0$ with
constant $G,\theta, G_s$ is
\be
\label{formallim}
{\bar g} \rightarrow -(2\pi\alpha')^2 \,{{\bar G} \over {\bar \theta}^2}  
,\qquad {\bar B} \rightarrow {1\over {\bar \theta}}, \qquad G_s \rightarrow
\sqrt{-g_s^2 \left({{\bar G}{\bar \theta} \over 2\pi\alpha'}\right)^2}.
\ee
In particular, the equation for the sigma-model metric components ${\bar g}$
 forces them to have opposite
 sign to those of the open-string metric ${\bar G}$. In addition, a real
Yang--Mills coupling or, equivalently, a real $G_s$, requires an analytic
continuation to imaginary values of the `closed-string' coupling $g_s$.
 Therefore, the formal scaling limit that induces the time-space 
noncommutative Moyal products is the Seiberg--Witten scaling limit, 
supplemented with two rather unconventional operations:

\begin{itemize}

\item An analytic continuation of the closed-string coupling into imaginary
values $g_s^2 \rightarrow -g_s^2$.

\item
A switch of space-time signature, or `tumbling' of light cones in the
time-space noncommutative plane.

\end{itemize}

These two features are strongly reminiscent of black-hole horizons. Namely,
the `tumbling' of light cones is a standard feature of horizons, and the
continuation to imaginary $g_s$ is related to the tachyonic character of
the D-brane in this region. It is tempting to interpret the $(g_{\mu\nu})$-frame
and the $(G_{\mu\nu})$-frame as living in opposite sides of a horizon, so that
a     tachyonic particle (as seen in the $(g_{\mu\nu})$-frame)
 can  escape through the horizon and emerge    
  on the other side as   a standard non-tachyonic particle with respect
to the  $(G_{\mu\nu})$-frame.  

Although complex values of the dilaton seem like a rather exotic feature,
 it is not completely  obvious that the combined action of the
two above operations  would yield an inconsistent theory, when reinterpreted
in the $G_{\mu\nu}$ space-time.   Equivalently, it is not obvious that
the open-string perturbation theory with time-space noncommutative phases
and real $G_s$ would be inconsistent.

At the level of the low-energy effective theory, this is a question of
whether  the  perturbation theory with
time-space noncommutative Moyal products defines a consistent $S$-matrix. 
            Apparently acausal effects were reported in \cite{teleo}
at the tree level, but similar effects show up in the Veneziano amplitude
for open strings. Therefore, these effects are not  necessarily fatal
to the quantum $S$-matrix.
 The most likely candidate for a smooth theory
would be ${\cal N}=4$ super-Yang--Mills, which is largely safe from
the infrared singularities of \cite{iruv}, owing to the improved ultraviolet
behaviour.  For the spatially noncommutative
 $U(N)$ theory we also have a candidate large-$N$ master field via the
AdS/CFT correspondence \cite{malda, gkpw}.  At
large values of the 't Hooft coupling $\lambda_{\rm YM}
 = g_{\rm YM}^2 N$, it is
given by the near-horizon  D3-brane backgrounds  in  type IIB string theory
with $B$-fields \cite{MR,HI}.

We can construct an AdS/CFT dual of the time-space noncommutative 
theory starting from the supergravity solution of $N$ D3-branes with
electric NS $B$-field backgrounds \cite{MR}.      
In the string frame:
\be
\label{metric}
ds^2 = {1\over \sqrt{H(r)}} \left( d{\bf y}^2 +f(r)(-dt^2 + dx^2)  \right) +
\sqrt{H(r)}\left(dr^2 + r^2 \,d\Omega_5^2 \right)
\ee
with 
\be
H(r)=1+\left({R\over r}\right)^4 ,\qquad
 {1\over f(r)} = -{{\rm sh}^2 \,(\alpha) 
\over H(r)} + {\rm ch}^2 \,(\alpha).
\ee
The $B$-field and dilaton profiles are
\be
{\bar b}=2\pi\alpha' {\bar B} = -{\rm th}\,(\alpha) \,{f(r)\over H(r)}, \qquad 
e^{2\phi} = g_s^2 \,f(r).
\ee
The charge radius satisfies $R^4 = 4\pi g_s N (\alpha')^2$.
   
The  zero-slope limit in eq. (\ref{formallim}) can be implemented by rescaling
the coordinates as
$$
(-dt^2 + dx^2) \rightarrow -\left({2\pi\alpha' \over {\bar \theta}}\right)^2
\,(-dt^2 + dx^2)
.$$
This induces a corresponding rescaling of ${\bar B}$ by a factor of
$-(2\pi\alpha' /{\bar \theta})^2$ (from the tensor transformation),
 whereas the analytic continuation
in $g_s$ induces 
$$
e^{2\phi} \rightarrow -e^{2\phi} 
.$$
 In order to satisfy
the $B$-field boundary condition in (\ref{formallim}) we must take 
${\rm th}\,(\alpha) \rightarrow \infty$. 
This can be achieved  by defining $t_\theta$ such that
$$
t_\theta \equiv {\rm th}\,(\alpha), \qquad {\rm sh}^2\,(\alpha) = {t_\theta^2
 \over 
1-t_\theta^2}, \qquad {\rm ch}^2 \,(\alpha) = {1\over 1-t_\theta^2}.
$$
and continuing $t_\theta$ outside the interval $[-1,1]$ as 
$$
t_\theta \rightarrow {\bar \theta \over 2\pi\alpha'}
.$$  

The complete scaling limit is   
$\alpha' \rightarrow 0$ with the previous prescriptions, plus the standard
AdS/CFT scaling $H\rightarrow 1/(Ru)^4$, with $u$  a sliding energy scale of
the strongly coupled  theory. 

The scaling limit defines the strong-coupling noncommutativity length
$a_\theta$:  
\be
\label{stncs}
(a_\theta)^4  = R^4 \,t_\theta^2 \rightarrow \lambda_{\rm YM} \,
\left({{\bar \theta}
\over 2\pi}\right)^2
,\ee
with $\lambda_{\rm YM} = g^2_{\rm YM} N = 4\pi G_s N$ the 't Hooft coupling
of the field theory.   

The final result is the dual background:
\be 
\label{dualdual}
{ds^2 \over R^2} = u^2 \left( d{\bf y}^2 + {\hat f}(u) (-dt^2 + dx^2)\right)
+ {du^2 \over u^2} + d\Omega_5^2
\ee
\be 
\label{dudu}
{\bar B} = {1\over {\bar \theta}} \,(a_\theta\,u)^4 \,{\hat f}(u), \qquad 
e^{2\phi} = \left({\lambda_{\rm YM} \over 4\pi N}\right)^2 \,{\hat f}(u) 
\ee 
where the nontrivial profile function is
\be
\label{cordero}
{\hat f}(u) = {1\over 1-(a_\theta\,u)^4}.
\ee
We see that this is just the analytic continuation of the purely spatial
noncommutative master-field metric of refs. \cite{MR,HI} 
under 
\be
\label{anal}
y\rightarrow it, \qquad \theta^{yx} \rightarrow i\,\theta^{tx} =i\,{\bar 
\theta},
\qquad   B_{yx} \rightarrow -i B_{tx}.
\ee 
This is natural, given the fact that
 the supergravity background is the effective description
induced by the 
 sum over planar diagrams. Since the analytic continuations (\ref{anal}) 
generate, diagram by diagram, 
 the time-space noncommutative perturbation theory, 
it is not surprising that the resulting master field is obtained by the
same analytic continuation. Turning the argument around, we can
say that   this
 serves as a nontrival consistency check of the zero-slope formulas in
(\ref{formallim}).

The large-$N$ master field encoded in eqs. (4.33-35)  has two important
physical properties: First, like the magnetic-$B$ counterpart,
 the geometry approaches the
standard $AdS_5 \times {\bf S}^5$ in the infrared $u\rightarrow 0$. This
suggests that indeed this theory shows no dangerous infrared singularities
of the type discussed in \cite{iruv}. 

 The second important property is
 the existence of a
naked singularity at $u=a_\theta^{-1}$, which is absent in
 the case of the magnetic
counterpart. 
The supergravity approximation breaks down badly at physical length scales of
the order of the strong-coupling noncommutativity scale $a_\theta$, with
the metric $B$-field and dilaton blowing up. In fact, the entire `ultraviolet' 
 region at $u>a_\theta^{-1}$ is hard to interpret, since the local value
of the closed-string coupling $e^{\phi(u)}$ becomes imaginary. Notice also
the `tumbling' of light cones in this region, in much the same fashion as
the prescription for the microscopic parameters in (\ref{formallim}).   

   Thus, the microscopic $g$-singularity  seems to reappear,  
 even after the zero-slope limit (\ref{formallim}) has been taken, disguised as
a threshold singularity in the low-energy large-$N$ master field. Therefore,
   there
are grounds to suspect that field theories with time-space noncommutativity
may be ultraviolet-inconsistent  at the quantum mechanical level,
 at least within
the approximations involved in the above derivation  (large-$N$ and large
't Hooft coupling).

It is interesting to consider the  behaviour of thermodynamic quantities
in the vicinity of the naked singularity. The near-extremal version of
 (\ref{dualdual}) can be obtained by the substitution $dt^2 \rightarrow
h(u)dt^2, \; du^2 \rightarrow du^2 /h(u)$, with $h(u)=1-(u_0/u)^4$ a
Schwarzschild-like profile function. As the horizon at $u=u_0$
 approaches the 
singularity, the thermodynamic functions scale with the parameter $\xi_0 =
1-(a_\theta u_0)^4$. In particular, the entropy and specific heat
 {\it vanish} as
$S \sim  C_V \sim (V/a_\theta^3) \,\sqrt{\xi_0}$ while the internal energy diverges
logarithmicaly $E\sim -(V/a_\theta^4) \,{\rm log}\,(\xi_0)$. 
On the other hand, the Hawking temperature diverges when the horizon reaches
the singularity: $T\sim 1/(a_\theta \sqrt{\xi_0})$.  Therefore, as long as
 $T \leqsim a_\theta^{-1}$,  the large-$N$
 Euclidean
path integral with fixed-temperature boundary conditions is not dominated
by the singular geometry. This would suggest
that the theory makes sense
 if defined with a cutoff $u<u_\Lambda < a_\theta^{-1}$
 that stays well below the noncommutativity energy  scale.   
On the other hand, the fact that the large-$N$
 extensive thermodynamic functions depend on $\theta$
 (through $a_{\theta}$) contradicts the perturbative argument
of \cite{bigsus}, namely Moyal phases should cancel in planar vacuum diagrams. 
The same is true of the formally continued $a_\theta \rightarrow ia_\theta$
 metrics that, though nonsingular, show other nonstandard thermodynamical
features, such as a maximum temperature and a branch of negative specific
heat (see refs. \cite{MR,us}). It is tempting to interpret these pathologies
as further evidence for the quantum
 inconsistency of the time-space noncommutative
field theories. In this case, the lack of a Hamiltonian formalism would
be responsible for the failure of the thermodynamical interpretation of
the  periodically identified Euclidean backgrounds.

The naked singularity does not protect itself within the supergravity
approximation. For example, one can calculate the static action of a
D3 probe brane in this background,
\be
\label{probe}
S_{\rm probe} = T_{{\rm D}3} \int e^{-\phi} \sqrt{-{\rm det}(g+2\pi\alpha' B)}
+ T_{{\rm D}3} \int C^{RR}_4 + T_{{\rm D}3} \int C^{RR}_2 \wedge 2\pi\alpha' B,
\ee
and find a flat potential as a function of the radial coordinate $u$. This
means that the singularity is reachable by Higgs expectation values in the
breaking of  the $U(N)$ group
into $U(1)$ factors.

We see two possible resolutions of this situation. Either the singularity
is a true one and signals an inconsistency of the time-space noncommutative
field theory, or the full type IIB closed-string
 theory resolves the singularity. In that
case we are presumably back into our `censorship' criterion: stringy fuzziness
is fully apparent at the time-space noncommutativity scale.

\vspace{1cm}

{\bf Note added:} While this paper was being prepared for publication,
there appeared  two
papers \cite{overlap} 
that had  substantial overlap with our results. In particular, these
articles propose a scaling of an  {\it interacting} theory
at the $g$-singularities, still with a time-space noncommutativity 
 parameter of the  order of the 
 string scale, as in (\ref{maxnc}). In
an S-dual description,  this is related to a purely space-space noncommutative
 Yang--Mills theory \cite{overlap, ganor}. The large-$N$ master fields turn
out to be the Lorentzian time-space noncommutative metrics in \cite{MR}. Since 
thermodynamic functions are invariant under  duality transformations, we see that
the
thermodynamics of the scaled theory at the $g$-singularity
  seems to be independent of $\theta$ to the leading
order in the large-$N$ limit.


\newpage 

\section*{\sc Acknowledgements}

J.L.F.B.    would like to thank Luis Alvarez-Gaum\'e, Giovanni Arcioni,
Joaquim Gomis, Karl Landsteiner, Esperanza L\'opez and
Miguel A. V\'azquez-Mozo for  
 many useful discussions on noncommutative field theory.  

E.R. would like to thank Shmuel Elitzur for useful discussions. 
The research of 
E.R. is  partially supported by the BSF-American--Israeli Bi-National
Science Foundation and the IRF Centres of Excellence Program.

%

\end{document}